\documentclass[12pt]{article}
\usepackage{amsfonts}
\usepackage[applemac]{inputenc}
\usepackage{amsmath,amssymb,fullpage}
\usepackage{makeidx}
\usepackage{graphicx}
\usepackage{color}
\usepackage{multicol}
\usepackage[bottom]{footmisc}
\usepackage{amsmath}
\usepackage{amssymb}
\providecommand{\U}[1]{\protect\rule{.1in}{.1in}}
\newcommand{\realio}{{\mathbb{R}_0}}

\newcommand{\fproof}{\hfill $\square$ \bigskip}
\newtheorem{example}{Example}[section]
\newtheorem{theorem}[example]{Theorem}
\newtheorem{definition}[example]{Definition}
\newtheorem{lemma}[example]{Lemma}
\newtheorem{problem}[example]{Problem}
\newtheorem{remark}[example]{Remark}
\newtheorem{corollary}[example]{Corollary}

\newcommand{\reali}{{\mathbb{R}}}

\numberwithin{equation}{section}

\def\EE{\mathbb{E}}

\def\cf{\mathcal{F}}

\def\1B{\text{1\!\!I}}

\begin{document}

\title{A financial market with singular drift and no arbitrage}
\author{Nacira AGRAM$^{1}$ and Bernt \O KSENDAL$^{2}$}
\date{21 August 2020}

\maketitle


\footnotetext[1]{Department of Mathematics, Linnaeus University SE-351\,95
V\"axj\"o, Sweden. \newline Email:\texttt{ nacira.agram@lnu.se.}}

\footnotetext[2]{Department of Mathematics, University of Oslo, Box 1053 Blindern, NO-0316 Oslo, Norway.
\newline Email: \texttt{oksendal@math.uio.no.}
\par
This research was carried out with support of the Norwegian Research Council,
within the research project Challenges in Stochastic Control, Information and
Applications (STOCONINF), project number 250768/F20.}

\begin{abstract}
We study a financial market where the risky asset is modelled by a geometric
It\^o-L\'{e}vy process, with a singular drift term. This can for example model a
situation where the asset price is partially controlled by a company which
intervenes when the price is reaching a certain lower barrier. See e.g. Jarrow
\& Protter \cite{JP} for an explanation and discussion of this model in the
Brownian motion case. As already pointed out by Karatzas \& Shreve \cite{KS}
(in the continuous setting), this allows for arbitrages in the market.
However, the situation in the case of jumps is not clear. Moreover, it is not
clear what happens if there is a delay in the system. \newline
In this paper we consider a jump diffusion market model with a singular drift
term modelled as the local time of a given process, and with a delay $\theta>
0$ in the information flow available for the trader. We allow the stock price dynamics to depend on both a continuous process (Brownian motion) and a jump process (Poisson random measure). We believe that jumps and delays are essential in order to get more realistic financial market models. Using white noise
calculus we compute explicitly the optimal consumption rate and portfolio in
this case and we show that the maximal value is finite as long as $\theta> 0$.
This implies that there is no arbitrage in the market in that case. However,
when $\theta$ goes to 0, the value goes to infinity. This is in agreement with
the above result that is an arbitrage when there is no delay.\newline
Our model is also relevant for high frequency trading issues. This is because high frequency trading often leads to intensive trading taking place on close to infinitesimal lengths of time, which in the limit corresponds to trading on time sets of measure 0. This may in turn lead to a singular drift in the pricing dynamics. See e.g.
Lachapelle \textit{et al} \cite{LLLL} and the references therein.\newline

\end{abstract}

\section{Introduction}

It is well-known that in the classical Black-Scholes market, there is no
arbitrage. However, if we include a singular term in the drift of the risky
asset, it was first proved by Karatzas \& Shreve \cite{KS} (Theorem B2, page 329), that arbitrages
exist. 
%See also El Karoui \cite{E}. 
Subsequently this type of market has been
studied by several authors, including Jarrow \& Protter \cite{JP}. They
explain how a singular term in the drift can model a situation where the asset
price is partially controlled by a large company which intervenes when the
price is reaching a certain lower barrier, in order to prevent it from going
below that barrier. They also prove that arbitrages can occur in such
situations.\newline The purpose of our paper is to extend this study in two directions:\\
First, we introduce jumps in the market. More precisely, we study a jump
diffusion market driven by a Brownian motion $B(\cdot)$ and an independent
compensated random measure $\widetilde{N}(\cdot,\cdot)$ with an added singular
drift term, modelled by a local time of an underlying L\'{e}vy process $Y(\cdot)$. 
In view of the unstable financial markets we have seen in recent years, and in particular during the economic crises in 2008 and the corona virus crisis this year, we think that jumps are useful in an attempt to obtain more realistic financial market models.\\
Introducing jumps in the stock price motion goes back to Cox \& Ross \cite{CR} and to Merton \cite{M}.\\
Second, we assume that the trader only has access to a \emph{delayed}
information flow, represented by the filtration $\mathcal{F}_{t-\theta}$,
where $\theta>0$ is the delay constant and $\mathcal{F}_{t}$ is the
sigma-algebra generated by both  $\{B(s)\}_{s\leq t}$ and $\{N(s,\cdot)\}_{s\leq t}$. 
This extension is also motivated by the effort to get more realistic market models. Indeed, in all real-life, markets there is delay in the information flow available, and traders are willing to pay to get the most recent price information. Especially, when trading with computers even fractions of seconds of delays are important.\newline
We represent the singular term by the local time of a given process and show
that as long as $\theta>0$ there is no arbitrage in this market. In fact, we
show that this delayed market is \emph{viable}, in the sense that the value of
the optimal portfolio problem with logarithmic utility is finite. However, if
the delay goes to $0$, the value of the portfolio goes to infinity, at least
under some additional assumptions.\newline We emphasize that our paper deals
with \emph{delayed information flow}, not delay in the coefficients in the
model, as for example in the paper by Arriojas \textit{et al} \cite{AHMP1}.
There are many papers on optimal stochastic control with delayed information
flow, also by us. However, to the best of our knowledge the current paper is
the first to discuss the effect of delay in the information flow on arbitrage
opportunities in markets with a singular drift coefficient. We will show that
by applying techniques from white noise theory we can obtain explicit
results. Specifically, our model is the following:\newline Suppose we
have a financial market with the following two investment possibilities:

\begin{itemize}
\item A risk free investment (e.g. a bond or a (safe) bank account), whose
unit price $S_{0}(t)$ at time $t$ is described by
\begin{align}
\begin{cases}
dS_{0}(t)=r(t)S_{0}(t)dt;\quad  t \in [0, T],\\
S_{0}(0)=1.\label{eq1.1a}
\end{cases}
\end{align}

\item A risky investment, whose unit price $S(t)$ at time $t$ is given by a
linear stochastic differential equation (SDE) of the form
\small{}
\begin{align}
\begin{cases}
dS(t) & =S(t^{-})[\mu(t)dt+\alpha(t)dL_{t}+\sigma(t)dB(t)+\int_{\mathbb{R}%
_{0}}\gamma(t,\zeta)\widetilde{N}(dt,d\zeta)]; \quad t \in [0,T],\\
S(0) & >0,\label{eq1.2a}
\end{cases}
\end{align}
\end{itemize}
where $\mathbb{R}_0=\mathbb{R} \setminus \{0\}.$
Here $B(\cdot)$ and $\widetilde{N}=N(dt,d\zeta)-\nu(d\zeta)dt$ is a standard
Brownian motion and an independent compensated Poisson random measure,
respectively, defined on a complete filtered probability space $(\Omega
,\mathcal{F},P)$ equipped with the filtration $\mathbb{F}=\{\mathcal{F}%
_{t}\}_{t\geq0}$ generated by the Brownian motion $B(\cdot)$ and $N(\cdot
)$. The measure $\nu$ is the L\'{e}vy measure of the Poisson random measure $N$,
and the singular term $L_{t}=L_{t}(y)$ is represented as the local time at a
point $y\in\mathbb{R}$ of a given $\mathbb{F}$-predictable process $Y(\cdot)$
of the form
\begin{equation}
Y(t)=\int_{0}^{t}\phi(s)dB(s)+\int_{0}^{t}\int_{\mathbb{R}_{0}}\psi
(s,\zeta)\tilde{N}(ds,d\zeta), \label{eq2.5}%
\end{equation}
for some real deterministic functions $\phi: [0,T] \rightarrow \mathbb{R},\psi:[0,T]\times \mathbb{R}_0\rightarrow \mathbb{R}$ satisfying
\begin{equation}
0<\int_{t}^{T}\Big\{\phi^{2}(t)+\int_{\mathbb{R}_{0}}\psi^{2}(t,\zeta
)\nu(d\zeta)\Big\}dt<\infty\text{ a.s. for all }t\in\lbrack0,T]. \label{eq2.4}%
\end{equation}
The coefficients $r(t),\mu(t),\alpha(t),$ $\sigma(t)>0$ and $\gamma
(t,\zeta)>0$ are given bounded $\mathbb{F}$-predictable processes, with
$\sigma(t)$ bounded away from $0$.\newline
In this market we introduce a \emph{portfolio process} $u:[0,T]\times
\Omega\rightarrow\mathbb{R}$ giving the fraction of the wealth 
%$X(t)$
 invested in the risky asset at time $t$, and a \emph{consumption rate process}
$c:[0,T]\times\Omega\rightarrow\mathbb{R}^{+}$ giving the fraction of the
wealth consumed at time $t$. We assume that at any time $t$ both $u(t)$ and
$c(t)$ are required to be adapted to a given possibly smaller filtration
$\mathbb{G}=\{\mathcal{G}_{t}\}_{t\in\lbrack0,T]}$ with $\mathcal{G}%
_{t}\subseteq\mathcal{F}_{t}$ for all $t$. For example, it could be a delayed
information flow, with
\begin{equation}
\mathcal{G}_{t}=\mathcal{F}_{\max(0,t-\theta)},\quad t\geq0,\text{ for some
delay }\theta>0. \label{delay}%
\end{equation}
This case will be discussed in detail later.\newline Let us denote by
$\mathcal{A}_{\mathbb{G}}$ the set of all admissible consumption and portfolio
processes. We say that $c$ and $u$ are admissible and write $c,u\in
\mathcal{A}_{\mathbb{G}}$ if, in addition, $u$ is self-financing and \small
$
\EE\Big[\int_{0}^{T}(u(t)^{2}+c(t)^{2})dt\Big]<\infty,
$
where $\EE$ denotes expectation with respect to $P$.
Note that if $c,u$ are admissible, then the corresponding wealth
process $X(t)=X^{c,u}(t)$ is described by the equation
\begin{align}
dX(t)  &  =X(t^{-})[(1-u(t))r(t)+u(t)\mu(t)-c(t)]dt+u(t)\alpha(t)dL_{t}%
\label{wealth}\\
&  +u(t)\sigma(t)dB(t)+u(t)\textstyle\int_{\mathbb{R}_{0}}\gamma
(t,\zeta)\widetilde{N}(dt,d\zeta)].\nonumber
\end{align}
For simplicity, we put the initial value $X(0)=1.$ \newline The optimal
consumption and portfolio problem we study is the following:

\begin{problem}
Let $a>0,b>0$ be given constants. Find admissible $c^{\ast},u^{\ast},$ such
that
\begin{equation}
J(c^{\ast},u^{\ast})=\sup_{c,u}J(c,u), \label{opt}%
\end{equation}
where
\begin{equation}
J(c,u)=\EE\Big[\int_{0}^{T}a\ln(c(t)X(t))dt+b\ln(X(T))\Big]. \label{j}%
\end{equation}
\end{problem}
\noindent Our results are the following:\\
Using methods from white noise calculus we find explicit expressions for
the optimal consumption rate $c^{*}(t)$ and the optimal portfolio $u^{*} (t)$.
Then we show that the value is finite for all positive delays in the
information flow. In particular, this shows that there is no arbitrage in that
case. This result appears to be new.\\
 We also show that, under additional assumptions, the value goes to
infinity when the delay goes to $0$. This shows in particular that also when
there are jumps the value is infinite when there is no delay, in agreement
with the arbitrage results of Karatzas \& Shreve \cite{KS} and Jarrow \&
Protter \cite{JP} in the Brownian motion case.

\begin{remark} 
In our problem we are using the logarithmic utility function, both for the consumption and for the terminal value.
It is natural to ask if similar results can be obtained for other utility functions.  The method used in this paper is quite specific for the logarithmic utility and will not work for other cases. This issue will be discussed in a broader context in a future research.
\end{remark}

\section{Preliminaries}
%\textcolor{red}{we did not refer to the Appendix}
As we have mentioned above, we will use white noise calculus to find explicit
expressions for the optimal consumption and the optimal portfolio.
Specifically, we will define the local time in the terms of the Donsker delta
function which is an element of the Hida space of stochastic distributions
$(\mathcal{S})^{\ast}$. A brief introduction to white noise calculus is given in the Appendix. For more information on the underlying white
noise theory we refer to Hida \textit{et al} \cite{HKPS}, Oliveira \cite{O},
Holden \textit{et al} \cite{HOUZ} and Di Nunno \textit{et al} \cite{DOP} and
Agram \& \O ksendal \cite{AO}.

\subsection{The Donsker delta function}

We now define the Donsker delta function and give some of its properties. It
will play a crucial role in our computations.

\begin{definition}
Let $Y:\Omega\rightarrow\mathbb{R}$ be a random variable which also belongs to
the Hida space $(\mathcal{S})^{\ast}$ of stochastic distributions. Then a
continuous functional
\begin{equation}
\delta_{Y}(\cdot):\mathbb{R}\rightarrow(\mathcal{S})^{\ast} \label{donsker}%
\end{equation}
is called a \emph{Donsker delta function} of $Y$ if it has the property that
\begin{equation}
\int_{\mathbb{R}}g(y)\delta_{Y}(y)dy=g(Y),\quad\text{a.s.}
\label{donsker property}%
\end{equation}
for all (measurable) $g:\mathbb{R}\rightarrow\mathbb{R},$ such that the
integral converges.
\end{definition}
\noindent Explicit formulas for the Donsker delta function are known in many cases. For
the Gaussian case, see Section 3.2. For details and more general cases, see
e.g. Aase \textit{et al} \cite{AaOU}.\\

In particular, for our process $Y$ described by the diffusion \eqref{eq2.5}, it is well known (see e.g.
\cite{MOP}, \cite{DiO}, \cite{DOP}) that the Donsker delta functional exists
in $(\mathcal{S})^{\ast}$ and is given by
\begin{align}
& \delta_{Y(t)}(y)=\frac{1}{2\pi}\textstyle{\int_{\mathbb{R}}}\exp^{\diamond
}\big[\textstyle{\int_{0}^{t}\int_{\mathbb{R}_{0}}}(e^{ix\psi(s,\zeta
)}-1)\tilde{N}(ds,d\zeta)+\int_{0}^{t}ix\phi(s)dB(s)\nonumber\label{eq2.7}\\
&  +\textstyle{\int_{0}^{t}\{ \int_{\mathbb{R}_{0}}}(e^{ix\psi(s,\zeta
)}-1-ix\psi(s,\zeta))\nu(d\zeta)-\frac{1}{2}x^{2}\phi^{2}(s)\} ds-ixy\big]dx,
\end{align}
where $\exp^{\diamond}$ denotes the Wick exponential.\\

Moreover, if $0\leq s\leq t,$ we can compute the conditional expectation
\begin{align}
\label{eq2.3} &  \mathbb{E}[\delta_{Y(t)}(y)|\mathcal{F}_{s}]\\
&  =\frac{1}{2\pi}\textstyle{\int_{\mathbb{R}}}\exp\big[\int_{0}^{s}%
\int_{\mathbb{R}_{0}}ix\psi(r,\zeta)\tilde{N}(dr,d\zeta)+\int_{0}^{s}%
ix\phi(r)dB(r)\nonumber\\
&  +\textstyle{\int_{s}^{t}\int_{\mathbb{R}_{0}}}(e^{ix\psi(r,\zeta)}%
-1-ix\psi(r,\zeta))\nu(d\zeta)dr-\int_{s}^{t}\frac{1}{2}x^{2}\phi
^{2}(r)dr-ixy\big]dx.\nonumber
\end{align}

Note that if we put $s=0$ in \eqref{eq2.3}, we get
\begin{align*}
\EE[\delta_{Y(t)}(y)]  &  =\textstyle{\frac{1}{2\pi}\int_{\mathbb{R}}%
\exp\Big(-\frac{1}{2}x^{2}\int_{0}^{t}\phi^{2}(r)dr}\\
&  +\textstyle{\int_{0}^{t}\int_{\mathbb{R}_{0}}(e^{ix\psi(r,\zeta)}%
-1-ix\psi(r,\zeta))\nu(d\zeta)dr-ixy\Big)dx}<\infty.
\end{align*}

Putting $\nu=0$ in \eqref{eq2.3}, yields
\begin{align}
&  \frac{1}{2\pi}\int_{\mathbb{R}}\exp\Big[\int_{0}^{s}%
ix\phi(r)dB(r)-\textstyle{\int_{s}^{t}\frac{1}{2}}x^{2}\phi^{2}%
(r)dr-ixy\Big]dx\nonumber\\
&  =\left(2\pi \textstyle{\int_{s}^{t}}\phi^{2}(r)dr\right)  ^{-\frac{1}{2}%
}\exp\Big(-\frac{\left(  \int_{0}^{s}\phi(r)dB(r)-y\right)  ^{2}}{2 \int_{s}%
^{t}\phi^{2}(r)dr}\Big), \label{eq1.7c}%
\end{align}
where we have used, in general, for $a>0,b\in\mathbb{R}$, that
\begin{equation}
\int_{\mathbb{R}}e^{-ax^{2}-2bx}dx=\sqrt{\frac{\pi}{a}}%
e^{\frac{b^{2}}{a}}. \label{eq2.12}%
\end{equation}

In particular, applying the above to the random variable $Y(t):=B_{y}(t)$ for some
$t\in(0,T]$ where $B_{y}$ is Brownian motion starting at $y$ , we get for all
$0\leq s < t$,
\begin{equation}\label{eq2.7}
\mathbb{E}[\delta_{B(t)}(y)|\mathcal{F}_s]=(2\pi (t-s))^{-\frac{1}{2}}\exp\Big[-\frac
{(B(s)-y)^{2}}{2(t-s)}\Big].
\end{equation}

We will also need the following estimate:

\begin{lemma}
Assume that $0\leq s\leq t\leq T$. Then
\begin{align}
\EE[\delta_{Y(t)}(y)|\mathcal{F}_{s}]\leq\Big(2\pi\textstyle{\int_{s}^{t}%
}\left\{  \phi^{2}(r)+\int_{\mathbb{R}_{0}}\psi^{2}(r,\zeta)\nu(d\zeta
)\right\}  dr \Big)^{-\frac{1}{2}}. \label{eq2.10}%
\end{align}

\end{lemma}

\noindent{Proof.} \quad From \eqref{eq2.3} we get, with $i=\sqrt{-1}$,
\begin{align*}
&  |\EE[\delta_{Y(t)}(y)|\mathcal{F}_{s}]|
 \leq\textstyle{\frac{1}{2\pi}\int_{\mathbb{R}}\exp}\left[  {\int_{s}%
^{t}\int_{\mathbb{R}_{0}}Re(e^{ix\psi(r,\zeta)}-1-ix\psi(r,\zeta))\nu
(d\zeta)dr-\frac{1}{2}\int_{s}^{t}x^{2}\phi^{2}(r)dr}\right]  {dx}\\
&  \leq\textstyle{\frac{1}{2\pi}\int_{\mathbb{R}}\exp}\left[  {\int_{s}%
^{t}\int_{\mathbb{R}_{0}}-\frac{1}{2}x^{2}\psi^{2}(r,\zeta)\nu(d\zeta
)dr-\frac{1}{2}\int_{s}^{t}x^{2}\phi^{2}(r)dr}\right]  {dx}\\
&  =\textstyle{\frac{1}{2\pi}\int_{\mathbb{R}}\exp}\left[  {-\frac{1}{2}%
x^{2}\int_{s}^{t}}\left\{  {\phi^{2}(r)+\int_{\mathbb{R}_{0}}\psi^{2}%
(r,\zeta))\nu(d\zeta)}\right\}  {dr}\right]  {dx}\\
&  =\textstyle{\Big(2\pi}\left(  {\int_{s}^{t}}\left\{  {\phi^{2}%
(r)+\int_{\mathbb{R}_{0}}\psi^{2}(r,\zeta)\nu(d\zeta)}\right\}  {dr}\right)
{\Big)^{-\frac{1}{2}}}.
\end{align*}
\fproof

%We also recall the following basic results:
%\begin{lemma}
%\begin{align}
%&\textstyle{E[\exp(\int_0^s ix\phi(r)dB_y(r))] = \exp(-\frac{1}{2} x^2 \int_0^s \phi^2(r)dr - ixy)},\\
%&\textstyle{E[\exp(\int_0^s \int_{\mathbb{R}}ix \psi(r,\zeta)\widetilde{N}(dr,d\zeta))]= \int_0^s \int_{\mathbb{R}} (e^{ix\psi(r,\zeta)} - 1 - ix\psi(r,\zeta)) \nu(d\zeta) dr},
%\end{align}
%where $B_y(\cdot)$ denotes Brownian motion starting at $y$.
%\end{lemma}

\subsection{Local time in terms of the Donsker delta function}

In this subsection we define the local time of $Y(\cdot)$ at $y$ and we give a
representation of it in terms of the Donsker delta function.
%First let us recall the definition of a particular version of the density of an occupation
%measure, which is referred to as a local time.

\begin{definition}
The local time $L_{t}(y)$ of $Y(\cdot)$ at the point $y$ and at time $t$ is
defined by
\[
L_{t}(y)=\lim_{\epsilon\rightarrow0}\frac{1}{2\epsilon}\lambda(\{s\in
\lbrack0,t];Y(s)\in(y-\epsilon,y+\epsilon)\}),
\]
where $\lambda$ denotes Lebesgue measure on $\mathbb{R}$ and the limit is in
$L^{2}(\lambda\times P)$.
\end{definition}

\begin{remark}
 Note that this definition differs from the definition in Protter \cite{P} Corollary 3, page 230, in two ways:\\
(i) We are using Lebesgue measure $d\lambda(s)=ds$ as integrator, not $d[Y,Y]_s$\\
(ii) Protter \cite{P} is defining left-sided and right-side local times. Our local time corresponds to the average of the two.\\
If the process $Y$ is Brownian motion both definitions coincide with the standard one.
We choose our definition because it is convenient for our purpose.
\end{remark}
\noindent There is a close connection between local time and the Donsker delta function
of $Y(t)$, given by the following result.

\begin{theorem}
The local time $L_{t}(y)$ of $Y$ at the point $y$ and the time $t$ is given
by the following $\mathcal{S}^{*}$-valued integral
\begin{equation}
L_{t}(y)=\int_{0}^{t}\delta_{Y(s)}(y)ds, \label{eq2.9}%
\end{equation}
where the integral converges in $(\mathcal{S})^{*}$.
\end{theorem}

\noindent{Proof.} \quad In the following we let $\chi_F$ denote the indicator function of the Borel set $F$, i.e.
\begin{align}
\chi_F(x)&= 
\begin{cases}
1 \text{ if } x \in F,\\
0 \text{ if } x \notin F.
\end{cases}
\end{align}
\noindent By definition of the local time and the Donsker delta
function, we have
\begin{align*}
L_{t}(y)  &  =\lim_{\epsilon\rightarrow0} \int_{0}^{t} \textstyle{\frac
{1}{2\epsilon}}\chi_{(y-\epsilon,y+\epsilon)} (Y(s))ds =\lim_{\epsilon
\rightarrow0} \int_{0}^{t} \Big(\int_{\mathbb{R}}\frac{1}{2\epsilon}%
\chi_{(y-\epsilon,y+\epsilon)} (x) \delta_{Y(s)}(x) dx\Big)ds\\
&  =\lim_{\epsilon\rightarrow0} \int_{\mathbb{R}}\textstyle{\frac{1}%
{2\epsilon}} \chi_{(y-\epsilon,y+\epsilon)} (x) \Big(\int_{0}^{t}
\delta_{Y(s)}(x) ds\Big)dx =\int_{0}^{t} \delta_{Y(s)}(y) ds,
\end{align*}
because the function $y \mapsto\delta_{Y(s)}(y)$ is continuous in
$(\mathcal{S})^{*}$.
%by \eqref{eq2.8a}.
\hfill$\square$ \bigskip

\section{Optimal consumption and portfolio in a market with a local time drift
term under partial information}

We now return to the model in the Introduction. Thus we consider the optimal
portfolio and consumption problem \eqref{opt}-\eqref{j} of an agent in the
financial market \eqref{eq1.1a} \& \eqref{eq1.2a}. The agent has access to a
partial information flow $\mathbb{G}=\{\mathcal{G}_{t}\}_{t\geq0}$ where
$\mathcal{G}_{t}\subseteq\mathcal{F}_{t}$ for all $t$. It is known
that if $\mathbb{G}=\mathbb{F}$, i.e. $\mathcal{G}_{t}=\mathcal{F}_{t}$ for
all $t$, and if there are no jumps ($N=\nu=0$), then the market is complete
and it allows an arbitrage. See Karatzas \& Shreve \cite{KS} and Jarrow \& Protter \cite{JP}. It is clear that our
market with jumps is not complete, even if $\mathbb{G}=\mathbb{F}$. However,
we will show that if $\mathcal{G}_{t}=\mathcal{F}_{t-\theta}$ for some delay
$\theta>0$, then the market is viable (i.e. the optimal consumption and
portfolio problem has a finite value) and it has no arbitrage. Moreover, we
will find explicitly the optimal consumption and portfolio rates. If the delay
goes to 0, we show that the value goes to infinity, in agreement with the
existence of arbitrage in the no-delay case.\newline First we need the
following auxiliary result.

\begin{lemma}
Suppose that $\EE[\delta_{Y(t)}(y)|\mathcal{G}_t] \in L^{2}(P)$ and that
\[
\mu(t)-r(t)+\alpha(t)\EE[\delta_{Y(t)}(y)|\mathcal{G}_{t}]>0.
\]
Then there exists a unique solution $u(t)=u^{\ast}(t)>0$ of the equation
\begin{align*}
&  (a+b)\sigma^{2}(t)u^{\ast}(t)+[a(T-t)+b]\int_{\mathbb{R}_{0}%
}\dfrac{u^{\ast}(t)\gamma^{2}(t,\zeta)}{1+u^{\ast}(t)\gamma(t,\zeta)}%
\nu(d\zeta)\\
&  =(a(T-t)+b)[\mu(t)-r(t)+\alpha(t)\EE[\delta_{Y(t)}(y)|\mathcal{G}_{t}]].
\end{align*}
\end{lemma}

\noindent{Proof.} \quad Define
\[
F(u)=a_{1}u+a_{2}\int_{\mathbb{R}_{0}}\frac{u\gamma^{2}(t,\zeta)}%
{1+u\gamma(t,\zeta)}\nu(d\zeta),\quad u\geq0,
\]
where $a_{1}=(a+b)\sigma^{2}(t),a_{2}=a(T-t)+b.$ Then
\[
F^{\prime}(u)=a_{1}+a_{2}\int_{\mathbb{R}_{0}}\frac{\gamma^{2}(t,\zeta
)}{(1+\gamma(t,\zeta))^{2}}\nu(d\zeta)>0,
\]
and
\[
F(0)=0,\quad\lim_{u\rightarrow\infty}F(u)=\infty.
\]
Therefore, for all $a>0$ there exists a unique $u>0$ such that $F(u)=a$.
\hfill$\square$ \bigskip\\
\noindent We can now proceed to our first main result:

\begin{theorem}
[Optimal consumption and portfolio]\label{thm3.3} Assume that
$\alpha$ and $\gamma>0$ are $\mathbb{G}$-adapted and that
\begin{align*}
&  \mathbb{E}[\delta_{Y(t)}(y)|\mathcal{G}_{t}]\in L^{2}(\lambda\times P)\text{ and }
\mathbb{E}[ \mu(t)-r(t)|\mathcal{G}_t] +\alpha(t)\mathbb{E}[\delta_{Y(t)}(y)|\mathcal{G}_{t}]>0,\text{ for all
}t\in\lbrack0,T].
\end{align*}
Then the optimal consumption rate is
\[
c^{\ast}(t)=c^{\ast}(t)=\frac{a}{b+a(T-t)},
\]
and the optimal portfolio is given as the unique solution $u^{\ast}(t)>0$ of
the equation
\begin{align*}
&  (a+b)\mathbb{E}[\sigma^{2}(t)|\mathcal{G}_t] u^{\ast}(t)+(a(T-t)+b)\int_{\mathbb{R}_{0}%
}\frac{u^{\ast}(t)\gamma^{2}(t,\zeta)}{1+u^{\ast}(t)\gamma(t,\zeta)}\nu
(d\zeta)\\
&  =(a(T-t)+b)\Big(\mathbb{E}[\mu(t)-r(t)|\mathcal{G}_t]+\alpha(t)\EE[\delta_{Y(t)}(y)|\mathcal{G}%
_{t}]\Big).
\end{align*}
In particular,
%if $\tilde{N}(t)=N(t)-\lambda t$, where $N(t)$ is a Poisson process with intensity $\lambda > 0$, we have $\nu(d\zeta)=\lambda \delta_{1}(d\zeta)$, where $\delta_{1}(d\zeta)$ is the point mass at 1 and
%\begin{align}
%\pi^{\ast}(t)  &  =\frac{(a(T-t)+b)[\mu(t)-r(t)+\alpha(t)E[\delta
%_{Y(t)}(y)|\mathcal{G}_{t}]]}{(a+b)\sigma^{2}(t)}.
%\end{align}
if there are no jumps ($N=\nu=0$), the optimal portfolio will be
\[
u^{\ast}(t)=\frac{(a(T-t)+b)\Big(\mathbb{E}[\mu(t)-r(t)|\mathcal{G}_t]+\alpha(t)\EE[\delta_{Y(t)}%
(y)|\mathcal{G}_{t}]\Big)}{(a+b)\mathbb{E}[\sigma^{2}(t)|\mathcal{G}_t]}.
\]

\end{theorem}

\noindent{Proof.} \quad By the It\^o formula for semimartingales, see e.g.
Protter \cite{P}, we get that the solution of \eqref{wealth} is
\begin{align*}
X(t)  &  =\exp\Big(\int_{0}^{t}u(s)\sigma(s)dB(s)+\int_{0}^{t}\left\{  r(s)+[\mu(s)-r(s)]u(s)-c(s)-\frac{1}{2}\sigma
^{2}(s)u^{2}(s)\right\}  ds\\
&  +\int_{0}^{t}u(s)\alpha(s)dL_{s}
+\int_{0}^{t}\int_{\mathbb{R}_{0}}\{\ln(1+u(s)\gamma(s,\zeta))-u(s)\gamma
(s,\zeta)\}\nu(d\zeta)ds\\
&  +\int_{0}^{t}\int_{\mathbb{R}_{0}}\ln\left\{  1+u(s)\gamma(s,\zeta
)\right\}  \widetilde{N}(ds,d\zeta)\Big).
\end{align*}
\noindent Since $\sigma$ and $\gamma$ are bounded and $u \in \mathcal{A}_{\mathbb{G}}$ the stochastic integrals in the exponent have expectation 0. Therefore we get
\begin{align}
\EE|\ln(X(t))]  &  =\EE\Big[\int_{0}^{t}\{r(s)+[\mu(s)-r(s)]u(s)-c(s)-\frac{1}%
{2}\sigma^{2}(s)u^{2}(s)\}ds\nonumber\\
&  +\int_{0}^{t}u(s)\alpha(s)dL_{s}+\int_{0}^{t}\int_{\mathbb{R}_{0}}%
\{\ln(1+u(s)\gamma(s,\zeta))-u(s)\gamma(s,\zeta)\}\nu(d\zeta)ds\Big].
\label{ln}%
\end{align}
Formulas \eqref{wealth} and \eqref{j} and the It\^o formula, lead to
\begin{align*}
&  J(c,u)=\EE\Big[\int_{0}^{T}a\ln(c(t)X(t))dt+b\ln(X(T))\Big]\\
&  =\EE\Big[\int_{0}^{T}\Big\{a\ln(c(t))+a\ln(X(t))+b\Big(r(t)+[\mu(t)-r(t)]u(t)-c(t)-\frac{1}{2}\sigma^{2}(t)u^{2}%
(t)\Big)\Big\}dt\\
&  +b\int_{0}^{T}u(t)\alpha(t)dL_{t}+b\int_{0}^{T}\int_{\mathbb{R}_{0}}\{\ln(1+u(t)\gamma(t,\zeta
))-u(t)\gamma(t,\zeta)\}\nu(d\zeta)dt\Big].
\end{align*}
Substituting \eqref{ln} in the above, gives
\begin{align*}
&  J(c,u)=\EE\Big[\int_{0}^{T}\Big\{a\ln(c(t))+a\Big(\int_{0}^{t}\{r(s)+[\mu(s)-r(s)]u(s)-c(s)-\frac{1}{2}\sigma
^{2}(s)u^{2}(s)\}ds\\
&  +\int_{0}^{t}u(s)\alpha(s)dL_{s}+\int_{0}^{t}\int_{\mathbb{R}_{0}}\{\ln(1+u(s)\gamma(s,\zeta))-\pi
(s)\gamma(s,\zeta)\}\nu(d\zeta)ds\Big)\\
&  +b\Big(\int_{0}^{T}\Big\{r(t)+[\mu(t)-r(t)]u(t)-c(t)-\frac{1}{2}\sigma
^{2}(t)u^{2}(t)\Big)\Big\}dt+\int_{0}^{T}u(t)\alpha(t)dL_{t}\\
&  +\int_{0}^{T}\int_{\mathbb{R}_{0}}\{\ln(1+u(t)\gamma(t,\zeta))-u(t)\gamma
(t,\zeta)\}\nu(d\zeta)dt\Big)\Big].
\end{align*}
Note that in general, we have, by the Fubini theorem,
\begin{align*}
\int_{0}^{T}\Big(\int_{0}^{t}h(s)ds\Big)dt  &  =\int_{0}^{T}\Big(\int_{s}%
^{T}h(s)dt\Big)ds=\int_{0}^{T}(T-s)h(s)ds=\int_{0}^{T}(T-t)h(t)dt,
\end{align*}
and%
\begin{align*}
\int_{0}^{T}\Big(\int_{0}^{t}h(s)dL_{s}\Big)dt  & =\int_{0}^{T}\Big(\int
_{s}^{T}h(s)dt\Big)dL_{s}=\int_{0}^{T}(T-s)h(s)dL_{s}=\int_{0}^{T}(T-t)h(t)dL_{t}.
\end{align*}
%and
%\begin{align*}
%\int_0^T \int_0^t (\int_{\mathbb{R}} h(s,\zeta) \nu(d\zeta))ds dt
%= \int_0^T \int_t^T \int_{\mathbb{R}} h(t,\zeta) \nu(d\zeta) dt
%\end{align*}
Therefore, using that
\[
dL_{t}=dL_{t}(y)=\delta_{Y(t)}(y)dt,
\]
we get from the above that
\begin{align}
&  J(c,u)=\EE\Big[\int_{0}^{T}E\Big[\Big\{a\Big(\ln(c(t))+(T-t)\{r(t)+[\mu
(t)-r(t)]u(t) -c(t)-\frac{1}{2}\sigma^{2}(t)u^{2}(t)\\
&  +(T-t)u(t)\alpha(t)\delta
_{Y(t)}(y)\nonumber\\
&  +(T-t)\int_{\mathbb{R}_{0}}\{\ln(1+u(t)\gamma(t,\zeta))-u(t)\gamma
(t,\zeta)\}\nu(d\zeta)\Big)dt\nonumber\\
&  +b\Big(r(t)+[\mu(t)-r(t)]u(t)-c(t)-\frac{1}{2}\sigma^{2}(t)u^{2}%
(t)\}+u(t)\alpha(t)\delta_{Y(t)}(0)\nonumber\\
&  +\int_{\mathbb{R}_{0}}\{\ln(1+u(t)\gamma(t,\zeta))-u(t)\gamma(t,\zeta
)\}\nu(d\zeta)\Big)\Big\}\Big|\mathcal{G}_{t}\Big]dt\Big].\nonumber
\end{align}
Using that $c,u,\alpha$ and $\gamma$ are $\mathbb{G}$-adapted, we obtain
\begin{align}
&  J(c,u)=\EE\Big[\int_{0}^{T}\Big\{a\Big(\ln(c(t))+(T-t)\{\mathbb{E}[r(t)|\mathcal{G}_t]+\mathbb{E}[\mu
(t)-r(t)|\mathcal{G}_t]u(t)\nonumber\\
&  -c(t)-\frac{1}{2}\mathbb{E}[\sigma^{2}(t)|\mathcal{G}_t]u^{2}(t)\}+(T-t)u(t)\alpha(t)\EE[\delta
_{Y(t)}(y)|\mathcal{G}_{t}]\nonumber\\
&  +(T-t)\int_{\mathbb{R}_{0}}\{\ln(1+u(t)\gamma(t,\zeta))-u(t)\gamma
(t,\zeta)\}\nu(d\zeta)\Big)\nonumber\\
&  +b\Big(\mathbb{E}[r(t)|\mathcal{G}_t]+\mathbb{E}[\mu(t)-r(t)|\mathcal{G}_t]u(t)-c(t)-\frac{1}{2}\mathbb{E}[\sigma^{2}(t)|\mathcal{G}_t]u^{2}%
(t)+u(t)\alpha(t)\EE[\delta_{Y(t)}(y)|\mathcal{G}_{t}]\nonumber\\
&  +\int_{\mathbb{R}_{0}}\{\ln(1+u(t)\gamma(t,\zeta))-u(t)\gamma(t,\zeta
)\}\nu(d\zeta)\Big)\Big\}dt\Big]. \label{eq3.3}%
\end{align}
This we can maximise pointwise over all possible values $c,u\in\mathcal{A}%
_{\mathbb{G}}$ by maximising for each $t$ the integrand. Then we get the
optimal consumption rate
\[
c^{\ast}(t)=\frac{a}{b+a(T-t)},
\]
and the optimal portfolio is given as the unique solution $u^{\ast}(t)>0$ of
the equation
\begin{align*}
&  (a+b)\mathbb{E}[\sigma^{2}(t)|\mathcal{G}_t]u^{\ast}(t)+[a(T-t)+b]\textstyle\int_{\mathbb{R}_{0}%
}\frac{u^{\ast}(t)\gamma^{2}(t,\zeta)}{1+u^{\ast}(t)\gamma(t,\zeta)}\nu
(d\zeta)\\
&  =(a(T-t)+b)\Big[\mathbb{E}[\mu(t)-r(t)|\mathcal{G}_t]+\alpha(t)\EE[\delta_{Y(t)}(y)|\mathcal{G}%
_{t}]\Big].
\end{align*}
In particular, if there are no jumps ($N=\nu=0$), we get
\[
u^{\ast}(t)=\frac{(a(T-t)+b)\Big[\mathbb{E}[\mu(t)-r(t)|\mathcal{G}_t]+\alpha(t)\EE[\delta_{Y(t)}%
(y)|\mathcal{G}_{t}]\Big]}{(a+b)\mathbb{E}[\sigma^{2}(t)|\mathcal{G}_t]}.
\]

\qquad\qquad\qquad\qquad\qquad\qquad\qquad\qquad\qquad\qquad\qquad\qquad
\qquad\qquad\qquad\qquad\qquad$\square$

\subsection{ The case when $\mathcal{G}_{t}=\mathcal{F}_{t-\theta}, \quad t
\geq0$}

From now on we restrict ourselves to the subfiltration $\mathcal{G}%
_{t}=\mathcal{F}_{t-\theta},t\geq0$ for some constant delay $\theta>0$, where
we put $\mathcal{F}_{t-\theta}=\mathcal{F}_{0}$ for $t\leq\theta$. In this
case we can compute the optimal portfolio and the optimal consumption
explicitly. By \eqref{eq2.3} we have the following result:

\begin{lemma}
Assume that $\alpha$ and $\gamma > 0$ are $\mathbb{G}_{\theta}$-adapted, where $\mathbb{G}_{\theta} =\{ \mathcal{F}_{t-\theta}\}_{t\geq 0}$. For $t \geq \theta$ we have
\begin{align}
& \EE [\delta_{Y(t)}(y)|\mathcal{F}_{t-\theta}]=\textstyle{\frac{1}{2\pi}%
\int_{\mathbb{R}}}\exp\big[\int_{0}^{t-\theta}\int_{\mathbb{R}_{0}}%
ix\psi(r,\zeta)\widetilde{N}(dr,d\zeta)+\int_{0}^{t-\theta}ix\phi
(r)dB(r)\nonumber\\
&  +\textstyle{\int_{t-\theta}^{t}}\int_{\mathbb{R}_{0}}(e^{ix\psi(r,\zeta
)}-1-ix\psi(r,\zeta))\nu(d\zeta)dr-\int_{t-\theta}^{t}\frac{1}{2}x^{2}\phi
^{2}(r)dr-ixy\big]dx. \label{eq3.15}%
\end{align}
In particular, if $\psi=0$ and $\phi=1$, we get $Y=B$ and (see also \eqref{eq2.7})
\begin{equation}
\EE[\delta_{B(t)}(y)|\mathcal{F}_{t-\theta}]=(2\pi\theta)^{-\frac{1}{2}}%
\exp\Big[-\frac{(B(t-\theta)-y)^{2}}{2\theta}\Big]. \label{eq3.16}
\end{equation}

\end{lemma}

\noindent Then by Theorem \ref{thm3.3}, we get

\begin{theorem}
\label{thm3.5} Suppose $\mathcal{G}_{t}=\mathcal{F}_{t-\theta}$ with
$\theta>0.$ Then the optimal consumption rate is given by
\[
c^{\ast}(t)=\frac{a}{b+a(T-t)},
\]
and the optimal portfolio is given as the unique solution $u^{\ast}(t)>0$ of the
equation
\begin{align*}
&  (a+b)\EE[\sigma^{2}(t)|\mathcal{F}_{t-\theta}]u^{\ast}(t)+(a(T-t)+b)\int_{\mathbb{R}_{0}%
}\frac{u^{\ast}(t)\gamma^{2}(t,\zeta)}{1+u^{\ast}(t)\gamma(t,\zeta)}\nu
(d\zeta)\\
&  =(a(T-t)+b)\big(\EE[\mu(t)-r(t)|\cf_{t-\theta}]+\alpha(t)\EE[\delta_{Y(t)}(y)|\mathcal{F}%
_{t-\theta}]\big).
\end{align*}
In particular,
\[
\sup_{c,u}J(c,u)=J(c^{\ast},u^{\ast})<\infty,
\]
and there is no arbitrage in the market.
\end{theorem}

\section{The limiting case when the delay goes to 0}

In this section we concentrate on the delay case and with optimal portfolio
only, i.e. without consumption. Thus we are only considering utility from terminal wealth, and we put $a=0$ and $b=1$ in Theorem \ref{thm3.3}.
 Moreover, we assume that $\phi=1$ and $\psi=0$, i.e. that
\begin{equation}
Y(t)=B(t); \quad t \in [0,T].
\end{equation}
\noindent Also, to simplify the calculations we assume that $r=0$ and $\mu
(t)=\mu>0,\alpha(t)=\alpha>0,\sigma(t)=\sigma>0$ are constants, and
$\gamma(t,\zeta)=\gamma(\zeta)$ is deterministic and does not depend on $t$. Then the wealth equation will take the form
\begin{align}
dX(t)  &  =X(t)[u(t)\mu dt+u(t)\alpha dL_{t}\label{wealth}\nonumber\\
&  +u(t)\sigma dB(t)+u(t)\textstyle\int_{\mathbb{R}_{0}}\gamma(\zeta
)\widetilde{N}(dt,d\zeta)];\quad t \in [0,T],\quad X(0)=1,
\end{align}
where the singular term $L_{t}=L_{t}(y)$ is represented as the local time at a
point $y\in\mathbb{R}$ of $B(\cdot)$. The performance functional becomes
\[
J_{0}(u)=\EE[\ln X^{(u)}(T)];\quad u\in\mathcal{A}_{\theta},
\]
where $\mathcal{A}_{\theta}$ denotes the set of all $\mathcal{F}_{t-\theta}$
-predictable control processes. This now gets the form
\begin{align}
&  J_{0}(u)=\EE\Big[\int_{0}^{T}\Big\{\mu u(t)-\frac{1}{2}\sigma^{2}%
u^{2}(t)+u(t)\alpha \EE[\delta_{Y(t)}(y)|\mathcal{F}_{t-\theta}]\nonumber\\
&  +\int_{\mathbb{R}_{0}}\{\ln(1+u(t)\gamma(\zeta))-u(t)\gamma(\zeta
)\}\nu(d\zeta)\Big\}dt\Big]. \label{eq3.3}%
\end{align}
\noindent Our second main result is the following:
\begin{theorem}
Suppose in addition to the above that 
\[
\int_{\mathbb{R}_{0}}\gamma^{2}(\zeta)\nu(d\zeta)<\sigma^{2}.
\]
Then
\[
\lim_{\theta\rightarrow0^{+}}\sup_{u\in\mathcal{A}_{\theta}}J_{0}(u)=\infty.
\]
In particular, if there is no delay ($\theta=0$) the value of the optimal
portfolio problem is infinite.
\end{theorem}

\noindent{Proof.} \quad For given $\theta>0$ choose
\[
u_{\theta}(t)=\frac{\mu+\alpha R}{\sigma^{2}},
\]
where we for simplicity put
\[
R=R_{\theta}=\EE[\delta_{B(t)}(y)|\mathcal{F}_{t-\theta}].
\]
Then we see that
\[
J_{0}(u_{\theta})\geq\frac{1}{2}\EE[(\mu+\alpha R)^{2}]\Big(1-{\frac
{\int_{\mathbb{R}_{0}}\gamma^{2}(\zeta)\nu(d\zeta)}{\sigma^{2}}}\Big)=:C_{1}%
\EE[(\mu+\alpha R)^{2}]\geq C_{2}+C_{3}\EE[R^{2}],
\]
since, by \eqref{eq2.10}, $\EE[R]<\infty$. Here $C_{1},C_{2},C_{3}$ are positive
constants.\newline It remains to prove that
\[
\EE[R_{\theta}^{2}]\rightarrow\infty\text{ when }\theta\rightarrow0^{+}.
\]
To this end, note that by \eqref{eq1.7c} we have
\begin{align}
\EE[R_{\theta}^{2}]=\EE[(\delta_{B(t)}(y)|\mathcal{F}_{t-\theta})^{2}]=(2\pi
\theta)^{-1}\EE\Big[\exp\Big(-\frac{2(B(t-\theta)-y)^{2}}{2\theta}\Big)\Big].
\end{align}
By formula 1.9.3(1) p.168 in \cite{BS} we have, with $\kappa>0$ constant,
\begin{align}
\label{BS}\EE[\exp(-\kappa(B(t-\theta)-y)^{2})]=\frac{1}{1+2\kappa(t-\theta)}
\exp\Big(-\frac{\kappa y^{2}(t-\theta)}{1+2\kappa(t-\theta)}\Big).
\end{align}

Applying this to $\kappa =\frac{1}{\theta}$ we get 
\begin{align}
\EE[R_{\theta} ^2]= \frac{1}{2\pi \sqrt{\theta} \sqrt{2t-\theta} } \exp\Big( -\frac{y^2}{2t-\theta} \Big)\rightarrow\infty,
\end{align}
when $\theta\rightarrow0$. \hfill$\square$

\section{The Brownian motion case}

In the case when $Y(t)=B(t)$ the computations above can be made more explicit.
We now illustrate this, assuming for simplicity that $y=0$. Then by Theorem
\ref{thm3.5} the optimal portfolio $\widehat{u}(t)$ is given by
\[
\widehat{u}(t)=\frac{\mu+\alpha\Lambda(t)}{\sigma^{2}},
\]
where
\begin{align}
\Lambda(t)  &  =\EE[\delta_{B(t)}(0)|\mathcal{F}_{t-\theta}]=(2\pi
\theta)^{-\frac{1}{2}}\exp\Big[-\frac{B(t-\theta)^{2}}{2\theta}\Big],\quad
t\geq\theta,\nonumber\\
&  \Lambda(t)=\frac{1}{\sqrt{2\pi\theta}},\quad0\leq t\leq\theta.
\label{eq4.1}%
\end{align}
By \eqref{eq3.3} and \eqref{eq4.1} we see, after some algebraic operations,
that the corresponding performance $\widehat{J}_{\theta}=J(0,\widehat{\pi})$
is
\begin{align*}
\widehat{J}_{\theta}  &  =\EE\Big[\int_{0}^{T}\Big(\mu\widehat{\pi}(t)-\frac
{1}{2}\sigma^{2}\widehat{\pi}^{2}(t)+\widehat{\pi}(t)\alpha\Lambda
(t)\Big)dt\Big]=\EE\Big[\int_{0}^{T}\frac{(\mu+\alpha\Lambda(t))^{2}}%
{2\sigma^{2}}dt\Big]\\
&  =A_{1}+A_{2}+A_{3},
\end{align*}
where
\[
A_{1}=\frac{\mu^{2}}{2\sigma^{2}}T,\quad A_{2}=\frac{\mu\alpha}{\sigma^{2}%
}E\Big[\int_{0}^{T}\Lambda(t)dt\Big],\quad A_{3}=\frac{\alpha^{2}}{2\sigma
^{2}}E\Big[\int_{0}^{T}\Lambda^{2}(t)dt\Big].
\]
Using the density of $B(s)$, we get
\begin{align}
\EE\Big[\exp\Big(-\frac{B^{2}(s)}{2\theta}\Big)\Big]  &  =\int_{\mathbb{R}}%
\exp\left(  -\frac{y^{2}}{2\theta}\right)  \frac{1}{\sqrt{2\pi s}}\exp\left(
-\frac{y^{2}}{2s}\right)  dy\nonumber\\
&  =\frac{1}{\sqrt{2\pi s}}\int_{\mathbb{R}}\exp\left(  -\frac{1}{2}%
y^{2}(\frac{1}{\theta}+\frac{1}{s})\right)  dy. \label{eq4.3}%
\end{align}
In general we have, for $a>0$,
\[
\int_{\mathbb{R}}\exp(-ay^{2})dy=\sqrt{\frac{\pi}{a}},
\]
we conclude, by putting $s=t-\theta$ in \eqref{eq4.3} , that
\[
A_{2}=\frac{\theta}{\sqrt{2\pi\theta}}+\int_{\theta}^{T}\frac{\mu\alpha
}{\sigma^{2}\sqrt{2\pi\theta}}\sqrt{\frac{\theta}{t}}dt=\sqrt{\frac{\theta
}{2\pi}}+\frac{2\mu\alpha(\sqrt{T}-\sqrt{\theta})}{\sigma^{2}\sqrt{2\pi}}.
\]
Finally we use similar calculations to compute
\[
A_{3}=\frac{\alpha^{2}}{2\sigma^{2}}(2\pi\theta)^{-1}\left(  \theta
+\int_{\theta}^{T}\Psi(t)dt\right)  ,
\]
where, putting $t-\theta=s$,
\begin{align*}
\psi(t)  &  =\EE\left[  \exp\left(  -\frac{B(s)^{2}}{\theta}\right)  \right]
=\int_{\mathbb{R}}e^{-\frac{y^{2}}{\theta}}\frac{1}{\sqrt{2\pi s}}%
e^{-\frac{y^{2}}{2s}}dy\\
&  =\frac{1}{\sqrt{2\pi s}}\int_{\mathbb{R}}\exp\left(  -y^{2}(\frac{1}%
{\theta}+\frac{1}{2s})\right)  dy=\frac{1}{\sqrt{2\pi s}}\sqrt{\frac{\pi
}{\frac{1}{\theta}+\frac{1}{2s}}}=\frac{1}{\sqrt{\frac{2s}{\theta}+1}}.
\end{align*}
This gives
\[
A_{3}=\frac{\alpha^{2}}{2\sigma^{2}}\left(  \frac{1}{2\pi}+\int_{0}^{T-\theta
}\frac{1}{2\pi\sqrt{\theta}\sqrt{2s+\theta}}ds\right)  =\frac{\alpha^{2}}%
{4\pi\sigma^{2}}\left(  1+\frac{\sqrt{2T-\theta}-\sqrt{\theta}}{\sqrt{\theta}%
}\right)  .
\]
We have proved the following:

\begin{theorem}
The optimal performance with a given delay $\theta> 0$ is given by\newline

$\widehat{J}_{\theta}=\frac{\mu^{2}}{2\sigma^{2}}T+\sqrt{\frac{\theta}{2\pi}%
}+\frac{2\mu\alpha(\sqrt{T}-\sqrt{\theta})}{\sigma^{2}\sqrt{2\pi}}%
+\frac{\alpha^{2}}{4\pi\sigma^{2}}(1+\frac{\sqrt{2T-\theta}-\sqrt{\theta}%
}{\sqrt{\theta}}).$\newline

In particular, $\widehat{J}_{\theta}\rightarrow\infty$ when $\theta
\rightarrow0.$
\end{theorem}

\begin{corollary}
(i) For all information delays $\theta>0$ the value of the optimal portfolio
problem is finite.\newline(ii) When there is no information delay, i.e. when
$\theta=0$, the value is infinite.
\end{corollary}

\section{Appendix}

In this section we give a brief survey of the underlying theory of
white noise analysis used in this paper. For more details see e.g.
Di Nunno \textit{et al} \cite{DOP} and Holden \textit{et al} \cite{HOUZ} and
the references therein.\newline

\begin{definition}
Let $\mathcal{S}(\mathbb{R})$ be the Schwartz space consisting of all
real-valued rapidly decreasing functions $f$ on $\mathbb{R},$ i.e.,
\begin{equation}
\lim_{|x|\rightarrow\infty}|x^{n}f^{(k)}(x)|=0,\quad\forall n,k\geq
0.\label{rapidely decreasing}%
\end{equation}

\end{definition}

\begin{example}
For instance $\mathcal{C}^{\infty}$ functions with compact support,
$f(x)=e^{-x^{2}},f(x)=e^{-x^{4}},...$ are all functions in $\mathcal{S}%
(\mathbb{R})$. 
\end{example}

\noindent For any $n,k\geq0,$ define a norm $\Vert.\Vert_{n,k}$ on $\mathcal{S}%
(\mathbb{R})$ by
\begin{equation}
\Vert f\Vert_{n,k}=\sup_{x\in\mathbb{R}}|x^{n}f^{(k)}(x)|.\label{norm on S}%
\end{equation}
Then the Schwartz space $\mathcal{S}(\mathbb{R)}$,
equipped with the topology defined by the family of seminorms \\
$\{\Vert.\Vert_{n,k},n,k\geq0\}$ is a Fr\' echet space.\newline

\noindent Let
$\mathcal{S}^{\prime}(\mathbb{R})$ be the dual space of $\mathcal{S}%
(\mathbb{R})$. $\mathcal{S}^{\prime}(\mathbb{R})$ is called the space of
tempered distributions. Let $\mathcal{B}$ denote the family of all Borel
subsets of $\mathcal{S}(\mathbb{R})$ equipped with the weak topology.\newline
From now on we will use the notation $\langle a,b\rangle$ that means $a$
acting on $b$.

\begin{theorem}
[Minlos] Let $\mathbf{E}$ be a Fr\' echet space with dual space $\mathbf{E}%
^{\ast}$. A complex-valued function $\phi$ on $\mathbf{E}$ is the
characteristic functional of a probability measure $\nu$ on $\mathbf{E}^{\ast
}$ ,i.e.,
\begin{equation}
\phi(y)=\int_{E^{\ast}}e^{i\langle x,y\rangle}d\nu(x),\quad y\in
\mathbf{E,}\label{Minlos}%
\end{equation}
if and only if it satisfies the following conditions:

\begin{enumerate}
\item $\phi(0)=1$,

\item $\phi$ is positive definite, i.e.
\[
\sum_{j,k=1}^{n}z_{j}\bar{z}_{k}\phi(a_{j}-a_{k})\geq0\text{ for all }%
z_{j},z_{k}\in\mathbb{C},a_{j},a_{k}\in\mathbf{E},
\]

\item $\phi$ is continuous.
\end{enumerate}
\end{theorem}

\begin{remark}
The measure $\nu$ is uniquely determined by $\phi$. Observe that $\phi
(0)=\nu(\mathbf{E}^{\ast}).$ Thus when condition $1$ above is not assumed, then we
can only conclude that $\nu$ is a finite measure.
\end{remark}
\subsection {White noise for Brownian motion}
\subsubsection {Construction of Brownian motion}
Let $\phi$ be the function on $\mathcal{S}(\mathbb{R})$ given by
\[
\phi(\xi)=\exp(-\frac{1}{2}|\xi|^{2}),\quad\xi\in\mathcal{S}(\mathbb{R}),
\]
where $|\cdot|$ is the $L^{2}(\mathbb{R})$ norm.\newline Then it is easy to
check that conditions 1-3 above are satisfied.\newline Therefore, by the
Minlos theorem there exists a unique probability measure $P$ on $\mathcal{S}%
^{\prime}(\mathbb{R})$ such that
\begin{equation}
\exp\left(  -\frac{1}{2}|\xi|^{2}\right)  =\int_{\mathcal{S}^{\prime
}(\mathbb{R})}e^{i\langle \omega,\xi\rangle}dP(\omega),\quad\xi\in\mathcal{S}%
(\mathbb{R}).\label{eq1.5}%
\end{equation}

\begin{definition}
The measure $P$ is called the standard Gaussian measure on $\mathcal{S}%
^{\prime}(\mathbb{R})$. The probability space $(\mathcal{S}^{\prime
}(\mathbb{R}),\mathcal{B},P)$ is called the white noise probability space. In
the following we will use the notation $\Omega=\mathcal{S}^{\prime}%
(\mathbb{R})$ and the elements of $\Omega$ are denoted by $\omega$.
The expectation with respect to $P$ is denoted by $\EE[\cdot]$,
\end{definition}

\noindent Note that from \eqref{eq1.5} it follows that
\begin{align}
\EE[\left\langle \omega,\xi\right\rangle ] &  =0\text{ for all }\xi
\in\mathcal{S}(\mathbb{R})\text{ and, }\label{eq1.7}\\
\EE[\left\langle \omega,\xi\right\rangle ^{2}] &  =|\xi|^{2}\text{ for all }%
\xi\in\mathcal{S}(\mathbb{R})\text{ (The Ito isometry).}\label{eq1.8}%
\end{align}
Using the Ito isometry we see that we can extend the definition of
$\left\langle \omega,\xi\right\rangle $ from $\xi\in\mathcal{S}(\mathbb{R})$
to all $\phi\in L^{2}(\mathbb{R})$ as follows:
\[
\left\langle \omega,\phi\right\rangle =\lim_{n\rightarrow\infty}\left\langle
\omega,\xi_{n}\right\rangle \text{ (limit in }L^{2}(P)),
\]
for any sequence $\xi_{n}\in\mathcal{S}(\mathbb{R})$ converging to $\phi$ in
$L^{2}(\mathbb{R}).$\newline Thus for each $t$ we can define $B(t,\cdot)\in
L^{2}(P)$ by
\[
B(t,\omega)=\left\langle \omega,\chi_{[0,t]}(\cdot\right\rangle ),\quad
t\geq0,\omega\in\Omega.
\]
Then the process $\{B(t,\omega)\}_{t\geq0,\omega\in\Omega}$ has stationary
independent increments of mean $0$ (by \eqref{eq1.7}), and the variance of
$B(t)$ is $t$ (by \eqref{eq1.8}). Moreover, by the Kolmogorov continuity
theorem the process has a continuous version. This version is a Brownian
motion. This is the Brownian motion we work with in this paper.

\subsubsection{The Wiener-It\^o chaos expansion}

Let the Hermite polynomials $h_{n}(x)$ be defined by
\[
h_{n}(x)=(-1)^{n}e^{\frac{1}{2}x^{2}}\frac{d^{n}}{dx^{n}}(e^{-\frac{1}{2}%
x^{2}}),\quad n=0,1,2,...
\]
The first Hermite polynomials are
\begin{equation}
h_{0}(x)=1,\quad h_{1}(x)=x,\quad h_{2}(x)=x^{2}-1,\quad h_{3}(x)=x^{3}%
-x,...\nonumber
\end{equation}
Let $e_{k}$ be the k$^{th}$ Hermite function defined by
\begin{equation}
e_{k}(x):=\pi^{-\frac{1}{4}}((k-1)!)^{-\frac{1}{2}}e^{-\frac{1}{2}x^{2}%
}h_{k-1}(\sqrt{2}x),\quad k=1,2,...\label{hermite function}%
\end{equation}
Then $\{e_{k}\}_{k\geq1}$ constitutes an orthonormal basis for $L^{2}(R)$ and
$e_{k}\in\mathcal{S}(\mathbb{R})$ for all $k$. Define
\begin{equation}
\theta_{k}(\omega):=\langle\omega,e_{k}\rangle=\int_{\mathbb{R}}%
e_{k}(x)dB(x,\omega),\quad\omega\in\Omega.\label{theta_k}%
\end{equation}
Let $\mathcal{J}$ denote the set of all finite multi-indices $\alpha
=(\alpha_{1},\alpha_{2},...,\alpha_{m}),m=1,2,...,$ of non-negative integers
$\alpha_{i}.$ If $\alpha=(\alpha_{1},...,\alpha_{m})\in\mathcal{J},\alpha
\neq0,$ we put
\begin{equation}
H_{\alpha}(\omega):=\prod_{j=1}^{m}h_{\alpha_{j}}(\theta_{j}(\omega
)),\quad\omega\in\Omega.\label{H_alpha}%
\end{equation}
By a result of It\^o we have that
\begin{equation}
I_{m}(e^{\widehat{\otimes}\alpha})=\prod_{j=1}h_{\alpha_{j}}(\theta
_{j})=H_{\alpha},\label{I_m}%
\end{equation}
where $I_{m}$ denotes the m-iterated It\^o integral, defined below.\newline We
set $H_{0}:=1$. Here and in the sequel the functions $e_{1},e_{2},...$ are
defined in (\ref{hermite function}) and $\otimes$ and $\widehat{\otimes}$
denote the tensor product and the symmetrized tensor product,
respectively.\newline The family $\{H_{\alpha}\}_{\alpha\in\mathcal{J}}$ is an
orthogonal basis for the Hilbert space $L^{2}(P)$. In fact, we have the
following result.

\begin{theorem}
[The Wiener-It\^o chaos expansion theorem (I)]\label{wiener ito1}The family
$\{H_{\alpha}\}\alpha\in\mathcal{J}$ constitutes an orthogonal basis of
$L^{2}(P)$. More precisely, for all $\mathcal{F}_{T}$-measurable $X\in
L^{2}(P)$ there exist (uniquely determined) numbers $c_{\alpha}\in\mathbb{R},$
such that
\begin{equation}
X=\sum_{\alpha\in\mathcal{J}}c_{\alpha}H_{\alpha}\in L^{2}%
(P).\label{decomposition1}%
\end{equation}
Moreover, we have
\begin{equation}
\Vert X\Vert_{L^{2}(P)}^{2}=\sum_{\alpha\in\mathcal{J}}\alpha!c_{\alpha}%
^{2}.\label{Norm1}%
\end{equation}

\end{theorem}

Let us compare the above Theorem to the equivalent formulation of this theorem
in terms of iterated It\^o integrals. In fact, if $\psi(t_{1},t_{2},...,t_{n})$
is a real deterministic symmetric function in its $n$ variables $t_{1}%
,...,t_{n}$ and $\psi\in L^{2}(\mathbb{R}^{n}),$ that is,
\[
\Vert\psi\Vert_{L^{2}(\mathbb{R}^{n})}:=\int_{\mathbb{R}^{n}}|\psi
(t_{1},t_{2},...,t_{n})|^{2}dt_{1}dt_{2}...dt_{n}
\]
then its $n$-iterated It\^o integral is defined by
\begin{align*}
&  I_{n}(\psi):=\int_{\mathbb{R}^{n}}\psi dB^{\otimes n}\\
&  =n!\int_{-\infty}^{\infty}\int_{-\infty}^{t_{n}}\int_{-\infty}^{t_{n-1}%
}...\int_{-\infty}^{t_{2}}\psi(t_{1},t_{2},...,t_{n})dB(t_{1})dB(t_{2}%
)...dB(t_{n}),
\end{align*}
where the integral on the right-hand side consists of $n$-iterated It\^ o integrals.
\newline 
Note that the integrand at each step is adapted to the
filtration $\mathbb{F}$. Applying the It\^o isometry $n$ times we see that
\begin{align}
\EE\big[\left(  \int_{\mathbb{R}^{n}}\psi dB^{\otimes n}\right)  ^{2}
\big]=n!\Vert\psi\Vert_{L^{2}(\mathbb{R}^{n})}^{2}.
\end{align}
For $n=0$ we adopt the convention that
\[
I_{0}(\psi):=\int_{\mathbb{R}^{0}}\psi dB^{\otimes0}=\psi=\Vert\psi
\Vert_{L^{2}(\mathbb{R}^{0})},
\]
for $\psi$ constant. Let $\widetilde{L}^{2}(\mathbb{R}^{n})$ denote the set of
symmetric real functions on $\mathbb{R}^{n}$, which are square integrable with
respect to Lebesgue measure.\newline Then we have the following result:

\begin{theorem}
[The Wiener It\^o chaos expansion theorem (II)]\label{wiener ito2}For all
$\mathcal{F}_{t}$- measurable $X\in L^{2}(P)$ there exist (uniquely
determined) deterministic functions $f_{n}\in\widetilde{L}^{2}(\mathbb{R}%
^{n})$ such that
\begin{equation}
X=\sum_{n=0}^{\infty}\int_{\mathbb{R}^{n}}f_{n}dB^{\otimes n}=\sum
_{n=0}^{\infty}I_{n}(f_{n})\in L^{2}(P).\label{decomposition2}%
\end{equation}
Moreover, we have the isometry
\begin{equation}
\Vert X\Vert_{L^{2}(P)}^{2}=\sum_{n=0}^{\infty}n!\Vert f_{n}\Vert
_{L^{2}(\mathbb{R}^{n})}^{2}.\label{isometry2}%
\end{equation}

\end{theorem}

The connection between these two expansions in Theorem (\ref{wiener ito1}) and
Theorem (\ref{wiener ito2}) is given by
\[
f_{n}=\sum_{\alpha\in\mathcal{J},|\alpha|=n}c_{\alpha}e_{1}^{\otimes\alpha
_{1}}\widehat{\otimes}e_{2}^{\otimes\alpha_{2}}\widehat{\otimes}%
...\widehat{\otimes}e_{m}^{\otimes\alpha_{m}},\quad n=0,1,2,...
\]
where $|\alpha|=\alpha_{1}+\alpha_{2}...+\alpha_{m}$ for $\alpha=(\alpha
_{1},...,\alpha_{m})\in\mathcal{J},m=1,2,...$ \vskip0.2cm Recall that the
functions $e_{1},e_{2},...$ are defined in (\ref{hermite function}) and
$\otimes$ and $\widehat{\otimes}$ denote the tensor product and the
symmetrized tensor product, respectively. \newline Note that since $H_{\alpha
}=I_{m}(e^{\widehat{\otimes}\alpha}),$ for $\alpha\in\mathcal{J},|\alpha|=m,$
we get that
\begin{equation}
m!\Vert e^{\widehat{\otimes}\alpha}\Vert_{L^{2}(\mathbb{R}^{m})}^{2}%
=\alpha!,\label{identit}%
\end{equation}
by combining (\ref{Norm1}) and (\ref{isometry2}) for $X=X_{\alpha}$.\newline

\subsubsection{Stochastic distribution spaces}

Analogous to the test functions $\mathcal{S}(\mathbb{R})$ and the tempered
distributions $\mathcal{S}^{\prime}(\mathbb{R})$ on the real line
$\mathbb{R},$ there is a useful space of stochastic test functions
$(\mathcal{S})$ and a space of stochastic distributions $(\mathcal{S})^{\ast}$
on the white noise probability space. We now explain this in detail:\newline In the
following we use the notation
\begin{equation}
(2\mathbb{N})^{\alpha}=\prod_{j=1}^{m}(2j)^{\alpha_{j}},\label{notation}%
\end{equation}
if $\alpha=(\alpha_{1},\alpha_{2},...)$.\newline We define
\[
\varepsilon^{(k)}=(0,0,...,1,...),
\]
with $1$ on the k$^{th}$ place. Thus we see that
\[
(2\mathbb{N})^{\varepsilon^{(k)}}=2k.
\]

\noindent \textbf{The Kondratiev Spaces $(\mathcal{S})_{1}, (\mathcal{S})_{-1}$
and the Hida Spaces $(\mathcal{S})$ and $(\mathcal{S})^{\ast}$}.

\begin{definition}
Let $\rho$ be a constant in $[0,1]$.
\end{definition}

\begin{itemize}
\item Let $k\in\mathbb{R}$. We say that $f=\sum_{\alpha\in\mathcal{J}%
}a_{\alpha}H_{\alpha}\in L^{2}(P)$ belongs to the Kondratiev test function
Hilbert space $(\mathcal{S})_{k,\rho}$ if
\begin{equation}
\Vert f\Vert_{k,\rho}^{2}:=\sum_{\alpha\in\mathcal{J}}a_{\alpha}^{2}%
(\alpha!)^{1+\rho}(2\mathbb{N})^{\alpha k}<\infty.\label{(S)_k norm}%
\end{equation}
%\end{definition}

\item We define the Kondratiev test function space $(\mathcal{S})_{\rho}$ as
the space
\[
(\mathcal{S})_{\rho}= \bigcap_{k\in\mathbb{R}}(\mathcal{S})_{k,\rho}%
\]
equipped with the projective topology, that is, $f_{n}\rightarrow f,
n\rightarrow\infty,$ in $(\mathcal{S})_{\rho}$ if and only if $\|f_{n}%
-f\|_{k,\rho}\rightarrow0,$ $n\rightarrow\infty,$ for all $k$.
%\end{definition}

\item Let $q\in\mathbb{R}$. We say that the formal sum $F = \sum_{\alpha
\in\mathcal{J}}b_{\alpha}H_{\alpha}$ belongs to the Kondratiev stochastic
distribution space $(\mathcal{S})_{-q,-\rho}$ if
\begin{equation}
\label{(S)_{-q} norm}\|f\|^{2}_{-q,-\rho}:=\sum_{\alpha\in\mathcal{J}%
}b_{\alpha}^{2}(\alpha!)^{1-\rho}(2\mathbb{N})^{-\alpha q}<\infty.
\end{equation}
We define the Kondratiev distribution space $(\mathcal{S})_{-\rho}$ by
%$(\mathcal{S})^{\ast}$ as the space%
\[
(\mathcal{S})_{-\rho} = \bigcup_{q\in\mathbb{R}}(\mathcal{S})_{-q,-\rho}%
\]
equipped with the inductive topology, that is, $F_{n}\rightarrow F,
n\rightarrow\infty,$ in $(\mathcal{S})_{-\rho}$ if and only if there exists
$q$ such that $\|F_{n}-F\|_{-q,-\rho}\rightarrow0, n\rightarrow\infty.$

\item If $\rho=0$ we write
\[
(\mathcal{S})_{0}=(\mathcal{S})\text{ and }(\mathcal{S})_{-0}=(\mathcal{S}%
)^{\ast}.
\]
These spaces are called the \emph{Hida test function space} and \emph{the Hida
distribution space}, respectively.
\end{itemize}

\begin{itemize}
\item If $F=\sum_{\alpha\in\mathcal{J}}b_{\alpha}H_{\alpha}$ in $(\mathcal{S}%
)_{-1}$, we define the generalized expectation $\EE[F]$ of $F$ by
\begin{equation}
\EE[F]=b_{0}.\label{expectation}%
\end{equation}
(Note that if $F\in L^{2}(P)$, then the generalized expectation coincides with
the usual expectation, since $\EE[H_{\alpha}]=0$ for all $\alpha\neq0$).\newline
\end{itemize}

Note that $(\mathcal{S})_{-1}$ is the dual of $(\mathcal{S})_{1}$ and
$(\mathcal{S})^{\ast}$ is the dual of $(\mathcal{S})$. The action of
$F=\sum_{\alpha\in\mathcal{J}}b_{\alpha}H_{\alpha}\in(\mathcal{S})_{-1}$ on
$f=\sum_{\alpha\in\mathcal{J}}a_{\alpha}H_{\alpha}\in(\mathcal{S})_{1}$ is
given by
\[
\langle F,f\rangle=\sum_{\alpha}\alpha!a_{\alpha}b_{\alpha}.
\]
We have the inclusion
\[
(\mathcal{S})_{1}\subset(\mathcal{S})\subset L^{2}(P)\subset(\mathcal{S}%
)^{\ast}\subset(\mathcal{S})_{-1}.
\]

\begin{example}
Since
\begin{align*}
B(t) &  =\left\langle \omega,\chi_{[0,t]}\right\rangle =\sum
_{k=1}^{\infty}(e_{k},\chi_{[0,t]})\left\langle \omega,e_{k}%
\right\rangle \\
&  =\sum_{k=1}^{\infty}\left(  \int_{0}^{t}e_{k}(s)ds\right)  H_{\varepsilon
^{(k)}},
\end{align*}
we see that \emph{white noise} $\overset{\bullet}{B}(t)$ defined by
\[
\overset{\bullet}{B}(t)=\frac{d}{dt}B(t)=\sum_{k=1}^{\infty}e_{k}%
(t)H_{\varepsilon^{(k)}},
\]
exists in $(\mathcal{S})^{\ast}$.
\end{example}

\subsubsection{The Wick product}

In addition to a canonical vector space structure, the spaces $(\mathcal{S})$
and $(\mathcal{S})^{\ast}$ also have a natural multiplication given by the
Wick product:

\begin{definition}
Let $X=\sum_{\alpha\in\mathcal{J}}a_{\alpha}H_{\alpha}$ and $Y=\sum_{\beta
\in\mathcal{J}}b_{\beta}H_{\beta}$ be two elements of $(\mathcal{S})^{\ast}$.
Then we define the Wick product of $X$ and $Y$ by
\[
X\diamond Y=\sum_{\alpha,\beta\in\mathcal{J}}a_{\alpha}b_{\beta}%
H_{\alpha+\beta}=\sum_{\gamma\in\mathcal{J}}\left(  \sum_{\alpha+\beta=\gamma
}a_{\alpha}b_{\beta}\right)  H_{\gamma}.
\]

\end{definition}

\begin{example}
We have
\[
B(t)\diamond B(t)=B^{2}(t)-t,
\]
and more generally
\begin{align*}
&  \left(  \int_{\mathbb{R}}\phi(s)dB(s)\right)  \diamond\left(
\int_{\mathbb{R}}\psi(s)dB(s)\right)  \\
&  =\left(  \int_{\mathbb{R}}\phi(s)dB(s)\right)  .\left(  \int_{\mathbb{R}%
}\psi(s)dB(s)\right)  -\int_{\mathbb{R}}\phi(s)\psi(s)ds,
\end{align*}
for all $\phi,\psi\in L^{2}(\mathbb{R})$.
\end{example}
\textbf{Some basic properties of the Wick product.} We list some properties of
the Wick product:

\begin{enumerate}
\item $X, Y\in(\mathcal{S})_{1} \Rightarrow X \diamond Y \in(\mathcal{S})_{1}$.

\item $X, Y\in(\mathcal{S})_{-1} \Rightarrow X \diamond Y \in(\mathcal{S}%
)_{-1}$.

\item $X, Y \in(\mathcal{S}) \Rightarrow X \diamond Y \in(\mathcal{S})$.

\item $X \diamond Y = Y \diamond X$.

\item $X \diamond(Y \diamond Z) = (X \diamond Y) \diamond Z$.

\item $X \diamond(Y + Z) = X \diamond Y + X \diamond Z$.

\item $I_{n}(f_{n})\diamond I_{m}(g_{m})=I_{n+m}(f_{n}\widehat{\otimes}%
g_{m}).$
\end{enumerate}

\noindent In view of the properties $(1)$ and $(4)$ we can define the Wick powers
$X^{\diamond n}$ $(n=1,2,...)$ of $X\in(\mathcal{S})_{-1}$ as
\[
X^{\diamond n}:=X\diamond X\diamond...\diamond X\text{ (n times) }.
\]
We put $X^{\diamond0}:=1$. Similarly, we define the Wick exponential
$\exp^{\diamond}X$ of $X\in(\mathcal{S})_{-1}$ by
\[
\exp^{\diamond}X:=\sum_{n=0}^{\infty}\frac{1}{n!}X^{\diamond n}\in
(\mathcal{S})_{-1}.
\]
In view of the aforementioned properties, we have that
\[
(X+Y)^{\diamond2}=X^{\diamond2}+2X\diamond Y+Y^{\diamond2},
\]
and also
\[
\exp^{\diamond}(X+Y)=\exp^{\diamond}X\diamond\exp^{\diamond}Y,
\]
for $X,Y\in\mathcal{S}_{-1}.$\newline 
Let $\EE[X]$ denote the generalized
expectation of an element $X \in (\mathcal{S})$. It coincides with the standard expectation if $X \in L^{1}(P)$. 
Then we see that
\[
\EE[X\diamond Y]=\EE[X]\EE[Y],
\]
for $X,Y\in(\mathcal{S})_{-1}$. Note that independence is not required for
this identity to hold. By induction, it follows that
\[
\EE[\exp^{\diamond}X]=\exp \EE{[X]},
\]
for $X\in(\mathcal{S})_{-1}$.

\subsubsection{Wick product, white noise and It\^o integration}

One of the spectacular results in white noise theory is the following, which
combines Wick product, white noise and It\^o integration:

\begin{theorem}
Let $\varphi(t) \in L^{2}([0,T] \times\Omega)$ be $\mathbb{F}$-adapted. Then
the integral $\int_{0}^{T} \varphi(t) \diamond\overset{\bullet}{B}(t) dt$
exists in $(\mathcal{S})^{*}$ and
\begin{equation}
\int_{0}^{T} \varphi(t) dB(t)= \int_{0}^{T} \varphi(t) \diamond\overset
{\bullet}{B}(t) dt.
\end{equation}

\end{theorem}

\begin{remark}
Heuristically, we can see that we obtain this result by using that
$\overset{\bullet}{B}(t)=\frac{d}{dt} B(t)$. If we work in $(\mathcal{S})^{*}$
this argument can be made rigorous.
\end{remark}
\subsection {White noise for  L\' evy process}
\subsubsection{Construction of  L\' evy processes} The construction we did above
for Brownian motion can be modified to apply to other processes. For
example, we obtain a white noise theory for L\' evy processes if we proceed
as follows (see \cite{DOP} for details):

\begin{definition}
Let $\nu $ be a measure on $\mathbb{R}_{0}$ such that 
\begin{equation}
\int_{\mathbb{R}}\zeta ^{2}\nu (d\zeta )<\infty .
\end{equation}%
Define 
\begin{equation}
h(\varphi )=\exp (\int_{\mathbb{R}}\Psi (\varphi (x))dx);\quad \varphi \in (%
\mathcal{S}),
\end{equation}%
where 
\begin{equation}
\Psi (w)=\int_{\mathbb{R}}(e^{iw\zeta }-1-iw\zeta )\nu (d\zeta );\quad w\in 
\mathbb{R},\quad i=\sqrt{-1}.
\end{equation}%
Then $h$ satisfies the conditions (i) - (iii) of the Minlos theorem 6.3. Therefore there exists a probability measure $%
Q $ on $\Omega =\mathcal{S}^{\prime }(\mathbb{R})$ such that 
\begin{equation}
\mathbb{E}_{Q}[e^{i\left\langle \omega ,\varphi \right\rangle
}]:=\int_{\Omega }e^{i\langle \omega ,\varphi \rangle}dQ(\omega )=h(\varphi );\quad
\varphi \in (\mathcal{S}).
\end{equation}%
The triple $(\Omega ,\mathcal{F},Q)$ is called the (pure jump) L\' evy
white noise probability space.
\end{definition}

One can now easily verify the following

\begin{itemize}
\item $\mathbb{E}_{Q}[\left\langle \cdot ,\varphi \right\rangle ]=0;\quad
\varphi \in (\mathcal{S})$

\item $\mathbb{E}_{Q}[\left\langle \cdot ,\varphi \right\rangle ^{2}]=K\int_{%
\mathbb{R}}\varphi ^{2}(y)dy;\quad \varphi \in (\mathcal{S})$, where $K=\int_{\mathbb{R}}\zeta ^{2}\nu (d\zeta ).$
\end{itemize}

\noindent As we did for the Brownian motion, we use an approximation argument to define 
\begin{equation}
\widetilde{\eta }(t)=\widetilde{\eta }(t,\omega )=\left\langle \omega ,\chi
_{\lbrack 0,t]}\right\rangle ;\quad a.a.(t,\omega )\in \lbrack 0,\infty
)\times \Omega .
\end{equation}%
Then the following holds:

\begin{theorem} \label{th5.2}
The stochastic process $\widetilde{\eta}(t) $ has a c\`adl\`ag version. This
version $\eta(t)$ is a pure jump L\' evy process with L\' evy measure $\nu$.
\end{theorem}

\subsubsection{Chaos expansion}
From now on we assume that the L\'{e}vy measure $\nu $ satisfies the
following condition:\\
For all $\varepsilon >0$ there exists $\lambda >0$ such that
\begin{equation}
\int_{\realio\backslash (-\varepsilon ,\varepsilon )}\exp (\lambda
\left\vert \zeta\right\vert )\nu (d\zeta)<\infty   \label{10.1}
\end{equation}
This condition implies that the polynomials are dense in $L^{2}(\rho )$,
where
\begin{equation}
\rho (d\zeta)=\zeta^{2}\nu (d\zeta)  \label{10.2}
\end{equation}
Now let $\left\{ l_{m}\right\} _{m\geq
0}=\left\{ 1,l_{1},l_{2},...\right\} $ \label{simb-1303}be the orthogonolization of 
$\left\{1,\zeta,\zeta^{2},...\right\} $ with respect to the inner product of $L^{2}(\rho )$.

Define
\begin{equation}
p_{j}(\zeta):=\left\Vert l_{j-1}\right\Vert _{L^{2}(\rho )}^{-1}\zeta_{j-1}(\zeta);
\text{ }j=1,2,...  \label{10.3}
\end{equation}
and
\begin{equation}
m_{2}:=\left( \int_\realio \zeta^{2}\nu (d\zeta)\right) ^{\frac{1}{2}}
=\left\Vert l_{0}\right\Vert _{L^{2}(\rho )}
=\left\Vert 1\right\Vert_{L^{2}(\rho )}.  \label{10.4}
\end{equation}
In particular,
\begin{equation}
p_{1}(\zeta)=m_{2}^{-1}\zeta\text{ or }\zeta=m_{2}p_{1}(\zeta).  \label{10.5}
\end{equation}
Then $\left\{ p_{j}(\zeta)\right\} _{j\geq 1}$ is an \emph{orthonormal basis}
for $L^{2}(\nu )$.

Define the bijection $\kappa :\mathbb{N}\times \mathbb{N}\longrightarrow
\mathbb{N}$\label{simb-new2} by
\begin{equation}
\kappa (i,j)=j+(i+j-2)(i+j-1)/2.  \label{10.6}
\end{equation}
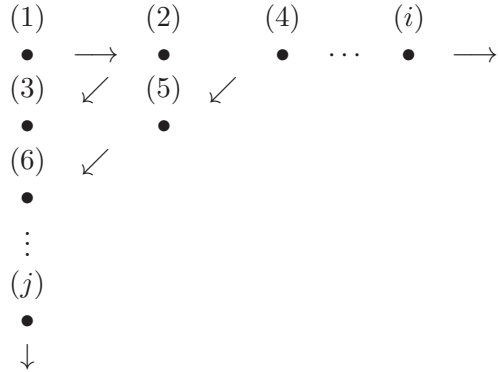
\[
\begin{array}{cccccccccc}
(1) &  & (2) &  & (4) &  & (i) &  &  &  \\
\bullet  & \longrightarrow  & \bullet  &  & \bullet  & \cdots  & \bullet  &
\longrightarrow  &  &  \\
(3) & \swarrow  & (5) & \swarrow  &  &  &  &  &  &  \\ 
\bullet  &  & \bullet  &  &  &  &  &  &  &  \\ 
(6) & \swarrow  &  &  &  &  &  &  &  &  \\ 
\bullet  &  &  &  &  &  &  &  &  &  \\ 
\vdots  &  &  &  &  &  &  &  &  &  \\
(j) &  &  &  &  &  &  &  &  &  \\ 
\bullet  &  &  &  &  &  &  &  &  &  \\ 
\downarrow  &  &  &  &  &  &  &  &  & 
\end{array}
\]
Let $\left\{ e _{i}(t)\right\} _{i\geq 1}$ be the Hermite functions.
Define
\begin{equation}
\delta _{\kappa (i,j)}(t,\zeta)=e _{i}(t)p_{j}(\zeta).  \label{10.7}
\end{equation}
If $\alpha \in \mathcal{J}$ with $Index(\alpha )=j$ and $\left\vert \alpha
\right\vert =m$, we define $\delta ^{\otimes \alpha }$ by
\begin{eqnarray}
&&\delta ^{\otimes \alpha }(t_{1},\zeta_{1},...,t_{m},\zeta_{m})  \label{10.8} \\
&=&\delta _{1}^{\otimes \alpha _{1}}\otimes ...\otimes \delta _{j}^{\otimes
\alpha _{j}}(t_{1},\zeta_{1},...,t_{m},\zeta_{m})  \nonumber \\
&=&\underset{\alpha _{1}\text{ factors}}{\underbrace{\delta
_{1}(t_{1},\zeta_{1})\cdot ...\cdot \delta _{1}(t_{\alpha _{1}},\zeta_{\alpha _{1}})}
}\cdot ...\cdot \underset{\alpha _{j}\text{ factors}}{\underbrace{\delta
_{j}(t_{m-\alpha _{j}+1},\zeta_{m-\alpha _{j}+1})\cdot ...\cdot \delta
_{j}(t_{m},\zeta_{m})}}.  \nonumber
\end{eqnarray}
We set $\delta _{i}^{\otimes 0}=1.$
Finally we let $\delta ^{\hat{\otimes }\alpha }$ denote the
\emph{symmetrized} tensor product\index{tensor product!symmetrized} of the $\delta _{k}$ $^{\prime }s:$
\begin{equation}
\delta ^{\hat{\otimes }\alpha }(t_{1},\zeta_{1},...,t_{m},\zeta_{m})
=\delta_{1}^{\hat{\otimes }\alpha _{1}}\otimes ...\otimes
\delta _{j}^{\hat{\otimes }\alpha _{j}}(t_{1},\zeta_{1},...,t_{m},z_{m}).  \label{10.9}
\end{equation}
For $\alpha \in \mathcal{J}$ define
\begin{equation}
K_{\alpha }:= I_{\left\vert \alpha \right\vert }\left(
\delta ^{\hat{\otimes }\alpha }\right) .  \label{10.10}
\end{equation}

\begin{theorem}
\label{Th10.2}
{\bf Chaos expansion.}\index{chaos expansion}
\newline Any $F\in L^{2}(P)$
has a unique expansion of the form
\begin{equation}
F=\sum_{\alpha \in \mathcal{J}}c_{\alpha }K_{\alpha }.
\label{10.14}
\end{equation}
with $c_{\alpha }\in \mathbb{R}$. 
Moreover,
\begin{equation}
\left\Vert F\right\Vert _{L^{2}(P)}^{2}=\sum_{\alpha \in \mathcal{J}}\alpha
!c_{\alpha }^{2}.  \label{10.15}
\end{equation}
\end{theorem}

\subsubsection{\bf The L\'{e}vy-Hida spaces} 
\begin{itemize}
\item[(i)] Let \ $(\mathcal{S})$ consist of all $%
\varphi =\sum_{\alpha \in \mathcal{J}}a_{\alpha }K_{\alpha }\in L^{2}(P)$
such that%
\begin{equation}
\left\Vert \varphi \right\Vert _{k}^{2}:=\sum_{\alpha \in \mathcal{J}%
}a_{\alpha }^{2}\alpha !(2\mathbb{N})^{k\alpha }<\infty \text{ for \emph{all}
}k\in \mathbb{N},  \label{10.18}
\end{equation}%
equipped with the projective topology, where%
\begin{equation}
(2\mathbb{N})^{k\alpha }=\prod_{j\geq 1}(2j)^{k\alpha _{j}},\text{ }
\label{10.19}
\end{equation}%
if $\alpha =(\alpha _{1,}\alpha _{2,},...)\in \mathcal{J}$.
\item[(ii)]
Let $(\mathcal{S})^{* }$ consist of all expansions $F=\sum_{\alpha \in \mathcal{%
J}}b_{\alpha }K_{\alpha }$ such that%
\begin{equation}
\left\Vert F\right\Vert _{-q}^{2}:=\sum_{\alpha \in \mathcal{J}}b_{\alpha
}^{2}\alpha !(2\mathbb{N})^{-q\alpha }<\infty \text{ for \emph{some} }q\in
\mathbb{N}.  \label{10.20}
\end{equation}%
endowed with the inductive topology.
The space $(\mathcal{S})^{* }$ is the
dual of $(\mathcal{S}).$ If $F=\sum_{\alpha \in \mathcal{J}}b_{\alpha
}K_{\alpha }\in (\mathcal{S})^{* }$ and $\varphi =\sum_{\alpha \in
\mathcal{J}}a_{\alpha }K_{\alpha }\in (\mathcal{S}),$ then the action of $F$
on $\varphi $ is%
\begin{equation}
\left\langle F,\varphi \right\rangle =\sum_{\alpha \in \mathcal{J}}a_{\alpha
}b_{\alpha }\alpha !.  \label{10.21}
\end{equation}
\item[(iii)]
If  $F= \sum_{\alpha \in \mathcal{J}} a_{\alpha }K_{\alpha } \in (\mathcal{S})^{* }$, we define
the generalized expectation $\EE[F]$ of $F$ by
$$
\EE[F] = a_0.
$$ 
Note that $\EE[K_{\alpha }] = 0$ for all $\alpha \ne 0$.
Therefore the generalized expectation coincides with the usual expectation if $F \in L^2(P)$. 
\end{itemize}

\bigskip\noindent
We can now define the white noise $\overset{\bullet }{\eta }(t)$\label{simb-1305} of the
L\'{e}vy process\index{white noise!of the L\'evy process}
\[
\eta (t)=\int_{0}^{t}\int_\realio \zeta\widetilde{N}(dt,d\zeta).
\]
and the white noise $\overset{\bullet }{\widetilde{N}}(t,\zeta)$ of $\widetilde{N}(dt,d\zeta)$ as follows.

\begin{equation}
\overset{\bullet }{\widetilde{N}}(t,\zeta)=\frac{\widetilde{N}(dt,d\zeta)}{dt\times
\nu (d\zeta)}\text{ (Radon-Nikodym derivative).}  \label{10.26}
\end{equation}%
Also note that $\overset{\bullet }{\eta }$ is related to $%
\overset{\bullet }{\widetilde{N}}$ by%
\begin{equation}
\overset{\bullet }{\eta }(t)=\int_\realio \overset{\bullet }{\widetilde{N%
}}(t,\zeta)\zeta\nu (d\zeta).  \label{10.27}
\end{equation}%

\subsubsection{The Wick product}

We now proceed as in the Brownian motion case and use the chaos expansion in
terms of $\left\{ K_{\alpha }\right\} _{\alpha \in \mathcal{J}}$ to define
the (L\'{e}vy-) Wick product.

\begin{definition}
\label{Def10.8} Let $F=\sum_{\alpha \in \mathcal{J}}a_{\alpha }K_{\alpha }$
and $G=\sum_{\beta \in \mathcal{J}}b_{\beta }K_{\beta }$ be two elements of $%
(\mathcal{S})^{* }.$ Then we define the \emph{Wick product} \index{Wick product}of $F$ and $G$
by%
\begin{equation}
F\diamond G=\sum_{\alpha ,\beta \in \mathcal{J}}a_{\alpha }b_{\beta
}K_{\alpha +\beta }=\sum_{\gamma \in \mathcal{J}}\left( \sum_{\alpha +\beta
=\gamma }a_{\alpha }b_{\beta }\right) K_{\gamma }.  \label{10.28}
\end{equation}
\end{definition}

\subsubsection{The Wick product, white noise and Skorohod integral}
%\textcolor {red}{can we have Levy-Ito integral not Skorohod one?}
\begin{theorem}
\label{Th10.11}
(i) Let $Y(t)$ be Skorohod integrable with respect to
$\eta .$ Then $Y(t)\diamond \overset{\bullet }{\eta }(t)$ is $dt-$integrable
in the space $(\mathcal{S})^{* }$ and%
\begin{equation}
\int_\reali Y(t)\delta \eta (t)=\int_\reali Y(t)\diamond \overset{%
\bullet }{\eta }(t)dt.  \label{10.34}
\end{equation}%
\newline(ii) Let $X(t,\zeta)$ be Skorohod-integrable with respect to $\widetilde{%
N}(\cdot ,\cdot ).$ Then $X(t,\zeta)\diamond \overset{\bullet }{\widetilde{N}}%
(t,\zeta)$ is $\nu (d\zeta)dt-$integrable in $(\mathcal{S})^{* }$ and%
\begin{equation}
\int_\reali \int_\realio X(t,\zeta)\widetilde{N}(\delta t,d\zeta)=\int_{\reali}
\int_\realio X(t,\zeta)\diamond \overset{\bullet }{\widetilde{N}}(t,\zeta)\nu
(d\zeta)dt.  \label{10.35}
\end{equation}
\end{theorem}

\section{Acknowledgments}

We are grateful to Martin Schweizer and an anonymous referee for helpful comments.

\end{document}